\documentclass[reqno, 11pt,a4paper]{amsart}
\usepackage{calc,graphicx,amsfonts,amsthm,amscd,epsfig,psfrag,amsmath,amssymb,enumerate,dsfont}
\usepackage{subfig} 
\usepackage{pdfsync}
\usepackage{stmaryrd} 
\usepackage[numeric,initials,nobysame]{amsrefs}

\usepackage[colorlinks=true, pdfstartview=FitV, linkcolor=blue, citecolor=blue, urlcolor=blue,pagebackref=false]{hyperref}

\usepackage[showlabels,sections,floats,textmath,displaymath]{}
\newcommand{\ie}{\hbox{\it i.e.\ }}

\DeclareMathSymbol{\leqslant}{\mathalpha}{AMSa}{"36} 
\DeclareMathSymbol{\geqslant}{\mathalpha}{AMSa}{"3E} 
\DeclareMathSymbol{\eset}{\mathalpha}{AMSb}{"3F}     
\renewcommand{\leq}{\;\leqslant\;}                   
\renewcommand{\geq}{\;\geqslant\;}                   

\renewcommand{\b}{\beta}

\renewcommand{\b}{\beta}
\renewcommand{\l}{\lambda}
\renewcommand{\L}{\Lambda}

\renewcommand{\l}{\lambda}
\renewcommand{\a}{\alpha}

\newcommand{\g}{\gamma}

\renewcommand{\O}{\Omega}

\newtheorem*{remark}{Remark}

\newtheorem*{conjecture}{Conjecture}

\let\a=\alpha \let\b=\beta     
 \let\g=\gamma       \let\l=\lambda
          \let\p=\pi  
  \let\s=\sigma    
  
     \let\L=\Lambda 
\let\O=\Omega      

\newcommand{\Tmix}{T_{\rm mix}}
\newcommand{\Trel}{ T_{\rm rel}} 
\newcommand{\Thit}{T_{\rm hit}}

\begin{document}
\author[P. Chleboun]{P. Chleboun}
 \address{P. Chleboun. Dip. Matematica,  Universit\`a  Roma Tre, Largo S.L.Murialdo 00146, Roma, Italy. e--mail:
paul@chleboun.co.uk}

\author{A. Faggionato}
\address{Alessandra Faggionato. Dip. Matematica ``G. Castelnuovo", Universit\`a ``La
  Sapienza''. P.le Aldo Moro  2, 00185  Roma, Italy. e--mail:
  faggiona@mat.uniroma1.it}

 \author[F. Martinelli]{F. Martinelli}
 \address{F. Martinelli. Dip. Matematica,  Universit\`a  Roma Tre, Largo S.L.Murialdo 00146, Roma, Italy. e--mail:
martin@mat.uniroma3.it
 }

\thanks{\sl Work supported by the European Research Council through the ``Advanced
Grant'' PTRELSS 228032.}

\title{Time Scale separation in the low temperature East model:
  Rigorous results}

\begin{abstract}
We consider the non-equilibrium
dynamics of the East model, a linear chain of $0$-$1$ spins evolving under a
simple Glauber dynamics in the
presence of a kinetic
constraint which forbids flips of those spins whose left neighbor is $1$. We focus on the glassy
effects caused by the kinetic constraint as
$q\downarrow 0$, where $q$ is the 
equilibrium density of the $0$'s. Specifically we analyse time scale
separation and dynamic
heterogeneity, i.e. non-trivial spatio-temporal fluctuations of the
local relaxation to equilibrium, one of the central aspects of glassy
dynamics. For any mesoscopic length scale $L=O(q^{-\g})$, $\g<1$, we
show that the characteristic time scale associated to two length scales
$d/q^\g$ and $d'/q^\g$ are
indeed separated by a factor $q^{-\a}$, $\a=\alpha(\g)>0$, provided
that $d'/d$ is large enough independently of $q$. In particular, the
evolution of 
mesoscopic \emph{domains}, {\it i.e.} maximal blocks of the form $111..10$,
occurs on a time scale which depends sharply on the size of the domain, a
clear signature of dynamic heterogeneity. Finally we show that \emph{no}
form of time scale separation can occur for $\g=1$, {\it
  i.e.} at the equilibrium scale
$L=1/q$, contrary to what was previously assumed in the physical
literature based on numerical simulations. 


\end{abstract}

\maketitle

\textbf{Introduction.} Kinetically constrained spins models (KCMs) are
stochastic particle models, usually  defined in terms of a non-interacting
Hamiltonian, whose dynamical evolution is determined by local rules encoding
a kinetic constraint. A typical move of the KCM's dynamics is the
creation/destruction of a particle with rates determined by the given
Hamiltonian through the usual detailed balance condition. However the move is inhibited unless the configuration of the
neighboring particles satisfies a certain constraint. 
Examples include the $f$-facilitated models \cite{FA2} which require
$f$ neighbors to be empty and anisotropic models in which certain
preassigned neighbors must be  empty.    

The main interest for these models (see e.g. 
\cites{Berthier,Ritort,SE2,GarrahanSollichToninelli,SE1,Cancrini:2006uu}) stems from the
fact  that KCMs, in spite of their simplicity, display many key dynamical
features of real glassy materials; ergodicity breaking
transition at some critical value $q_c$ of the vacancy density $q$, huge relaxation time for $q$
close to $q_c$, super-Arrehenius behaviour, dynamic heterogeneity
({\it i.e.} non-trivial spatio-temporal fluctuations of the
local relaxation to equilibrium) and aging, just to mention a
few. Mathematically, KCMs pose
very challenging and interesting problems because of the hardness of
the constraint, with ramifications towards bootstrap percolation
problems \cite{Spiral}, combinatorics \cites{CDG},
coalescence processes \cites{FMRT-cmp,FMRT} and  random walks on
triangular matrices \cite{Peres-Sly}. Quite surprisingly, some of the tools
developed for the analysis of the relaxation process of KCMs proved to
be quite powerful also in other contexts such as card shuffling problems
\cite{Bhatnagar:2007tr} and random evolution of surfaces
\cite{PietroCaputo:2012vl}.

In this paper we report on some rigorous findings \cite{CFM} concerning the glassy dynamics of the
East model, a popular KCM (see e.g. \cites{JACKLE,SE1,SE2} and
\cites{Aldous,CMRT,East-Rassegna}) defined on a one dimensional integer lattice of $L$ sites,
$\Lambda = \{1,2,\ldots,L\}$, for which each site can be in state 
$0$ or $1$, corresponding to \emph{empty} or \emph{occupied}
respectively. We denote the state space by $\O_\L:=\{0,1\}^\L$. The configuration evolves under Glauber type dynamics in
the presence of the kinetic constraint which forbids flips of those
spins whose left neighbour is in state $1$.
Each vertex, $x$, waits an independent mean one exponential time and then,
provided that the current configuration $\s$ satisfies the constraint
$\s_{x-1}=0$, the value of $\s$ at $x$ is refreshed and set equal to
$1$ with probability $1-q$ and to $0$ with probability $q$. The
leftmost vertex, $x = 1$, is unconstrained, equivalently we could
imagine that at $x=0$ there is  a frozen $0$ spin.
Since the constraint at site $x$ does not depend on the configuration
at site $x$, detailed balance is satisfied with respect to the product
Bernoulli measure $\pi:=\prod_{x\in\L}\p_x$, where $\pi_x$ is the Bernoulli
measure with density $p=1-q$.

 We focus on case $q\ll1$ because this is the regime where the
glassy features of the time evolution are more pronounced. It is also the situation which
requires more theoretical work, rigorous and non-rigorous, because of the huge relaxation times
which make numerical simulation unfeasible.   The equilibrium density
of zeros, $q$, is often written as $q =
\frac{e^{-\beta}}{1+e^{-\beta}}$, where $\beta$ is the inverse
temperature, so small $q$  corresponds to the low temperature regime.

Before describing our results we first briefly review
some previous work. It is known rigorously that the stationary dynamics relax exponentially fast on
a time scale $T_{\rm rel}$ which is uniformly bounded in the length $L$
of the chain
\cite{Aldous,CMRT} and diverges very rapidly  as $q\downarrow
0$ (cf. \eqref{piscina} below). It is also known \cite{CMST} that,
after time $T_{\rm rel}$ and starting from a large class of initial
laws, the local vacancy  density converges exponentially fast to $q$.

However the most interesting and
challenging dynamical behavior occurs for $q\ll1$ on time scales \emph{shorter} than the
relaxation time, $T_{\rm rel}$. It was first argued  in \cite{SE1,SE2}
and then later proved  in \cite{FMRT-cmp} that the dynamics of the infinite East chain
in a space window $[1,2^N]$ and up to time scales
$O(1/q^N)$\footnote{Recall that $f=O(g)$ means that $\exists\ c>0$
  independent of $q$
  such that $|f|\le c|g|$.} is
well approximated, as $q\downarrow 0$, by a certain \emph{hierarchical
  coalescence process} (HCP) \cite{FMRT}. In this HCP vacancies are
isolated and \emph{domains}, {\it i.e.}
maximal intervals of the form $111..10$, with cardinality
between $2^{n-1}+1$ and $2^n$, $n\le N$, are able to
coalesce with the domain on their right  only at time
scale $\sim (1/q)^n$.
As a result, aging and dynamic heterogeneity in the above regime emerge in a natural way, with a scaling limit\footnote{We emphasize that the physical literature
  (see e.g. \cites{SE1,Derrida}) assumed the existence of a scaling
limit. One of the main contribution of \cite{FMRT} was to actually
prove it together with a complete classification of the universality classes.} for the relevant quantities in the same universality class as several other mean field coalescence models of statistical physics \cite{Derrida}.

The  above result says nothing about the
dynamics and its characteristic time scales at
intermediate (mesoscopic) length scales $L=1/q^\g$, $0<\g<1$, or at the typical
inter-vacancy equilibrium scale $1/q$. At these length scales the low temperature dynamics
of the East model is no longer predominantly driven by an effective energy landscape, but entropic effects become crucial and
a subtle entropy/energy competition comes into play. As an example of
these effects, we observe that the natural
extrapolation to the
equilibrium scale $L_c=1/q$ of the
characteristic time scale $T_{\rm rel}^{(n)}\approx 1/q^n$, appropriate for domains of
length $L_n\approx 2^n$ with  $n=O(1)$, led (for example see
\cite{SE1}) to the wrong prediction of a relaxation time $\sim
2^{\theta^2}$, where $\theta=\log_2 (1/q)$, to be
compared with the correct scaling $2^{\frac{\theta^2}{2}(1+o(1))}$ (see
\eqref{piscina2} with $\g=1$). Notice that for small temperatures $\theta \simeq  \b/\ln(2)$.
\\

\noindent
\textbf{Results.}
Our first result is to show that, in the low temperature limit,
for a given system of size $L=O(1/q)$, three natural characteristic time
scales of the East Model are equivalent (up to universal constants). 
This equivalence is important because of various notions of
``equilibration time''  which appear in the physics literature.
The three characteristic times considered, on a system of size $L$ are
as follows:\\
(i) The relaxation time $\Trel(L)$, defined as the inverse of the spectral gap of the
infinitesimal generator of the process, \ie the worst relaxation time
found by fully diagonalising the master equation.
\\ 
(ii) The mixing time, $\Tmix(L)$, which is the time required for the
chain to be close
(in the sense of total variation distance) to the equilibrium
distribution $\pi$, starting from the worst case initial distribution.\\
(iii) The mean first passage time $\Thit(L)$; starting from the state $\sigma =
(1,1,\ldots,1,0)$ with a single vacancy at site $L$, we define
$\Thit(L)$ as the mean time for the spin at site $L$ to flip to $1$.
As observed in \cite{SE1}, the time $\Thit(L)$ gives some insight into the low
temperature dynamics, because of the
following argument. For small $q$ the process quickly reaches a state
in which there are few isolated $0$'s separated by \emph{domains} of $1$'s.
If two vacancies are separated by a domain  of cardinality $\ell$
then, conditioned on the vacancy on the left surviving, the mean time to
remove the vacancy on the right is exactly $\Thit(\ell)$. 

For a given $L=O(1/q)$, we establish that the ratio of any two of the
above time scales is bounded by a universal constant.

Having established the above equivalence, our next result
concerns the dependence of the characteristic time scales on the length $L$. In
particular we address the issue of \emph{time scale separation} up to the equilibrium scale $1/q$. 
For two system sizes $L',L$ which do not depend on $q$, $T_{\rm rel} \left( L' \right) \gg
T_{\rm rel} \left( L\right)$ if and only if $\lceil \log_2 L' \rceil >
\lceil \log_2 L \rceil$\footnote{Recall that $\lceil x\rceil$ is the
  smallest integer $n$ such that $n\ge x$.} (where $f \gg g$ means $f$ dominates $g$ by a
factor of $1/q^\a$, $\a>0$, as $q\downarrow 0$).
For mesoscopic length scales, that is for $L=d/q^\g$ for some $d>0$
and $\g \in (0,1)$, we show that there exists some $\l>1$ depending only
on $\g$ such that $T_{\rm rel} \left( \l L\right) \gg
T_{\rm rel} \left( L\right)$ . 
Moreover, if $ \gamma < 1/2$ then $\l = 2$. 

While at finite lengths (independent of
  $q$)  the question of time scale separation is completely
characterised, for mesoscopic lengths of order $O(1/q^\g)$, for
$\gamma \in (0,1)$, our knowledge is less detailed. Although we show that
time scale separation occurs between length scales whose ratio
is above a certain threshold $\l$, we do not know,
for example, 
if it occurs in a ``continuous'' fashion.
Given $\g\in (0,1]$ we say that \emph{continuous time scale separation}
occurs at length scale $1/q^\g$ if  $T_{\rm rel} \left( d'/ q^\gamma
\right) \gg T_{\rm rel} \left( d/q^\gamma \right)$ whenever $d'>d$.
This still remains an interesting open problem on mesoscopic length scales.

In \cite{SE1,SE2} a continuous time scale separation was conjectured
for $\g=1$, {\it i.e.} for length scales of the order of the
equilibrium inter-vacancy distance. The following hypothesis was
put forward on the basis of numerical simulations,
\begin{align}
\label{eq:cts:hyp}
 \left(\frac{  T_{\rm hit} (  d/q  ) }{ T_{\rm
      rel} ( \infty ) }\right)^{\frac{1}{\log(1/q) }} \to f(d) \quad
\textrm{as} \quad q \downarrow 0 \,,
\end{align}
for some positive, strictly increasing function $f(d)$.
This hypothesis was fundamental to the so-called
\emph{super-domain dynamics} proposed by Evans
and Sollich to describe the time evolution of the stationary East
model \cite{SE1,SE2}.
Notice that, because of the equivalence of the time scales $T_{\rm hit}(L), T_{\rm
  rel}(L)$ for $L=O(1/q)$, we could have replaced $T_{\rm hit}(d/q)$ in \eqref{eq:cts:hyp} by $T_{\rm rel}(d/q)$ 
without changing the scaling function $f$.  

Our next result shows that there is \emph{no form} of time scale separation
at the equilibrium length scale, that is for $\g = 1$, in particular
the above conjecture does not hold. In fact 
we are able to show that, as $q\downarrow 0$,
\begin{equation}
  \label{eq:3}
T_{\rm rel}(d/q)/T_{\rm rel}(d'/q)\le c(d,d'),  
\end{equation}
with $c(d',d)$ independent of $q$. So the ratio of the characteristic
time scales for system sizes which differ on the equilibrium length
does not diverge as $q\downarrow 0$, in particular this implies that
$f$ in \eqref{eq:cts:hyp} is constant. 
This is an important example of a case in which
numerical simulations can be misleading on systems where the
characteristic times are so large as to not be practically accessible
through simulations, and emphasizes the need
for rigorous results. 

Finally, building on these results,  we are able to prove
\cite{CFM} that the East process exhibits dynamic heterogeneity on
mesoscopic length scales, $1/q^\g$ for $\g \in (0,1)$, in the
following sense.  For a fixed $\g<1$,
starting from any initial configuration, after a time
$\Trel(1/q^\gamma)$ domains much shorter than $1/q^\g$ are very
unlikely to be present, whereas any domain that was much
larger than $1/q^\g$ initially is very likely to have survived. 

A key result required to prove the findings outlined above  was to establish 
extremely precise bounds on the relaxation time as a function of
$L$ and $q$. We prove that for any $d>0$ and $L \leq d/q$, there
exist positive exponents $\a,\a' $ depending only on  $d$ such that
\begin{equation}\label{piscina}
  \frac{n!}{q^n 2^{\binom{n}{2} }}q^{\a} \leq T_{\rm rel } (L) \leq
  \frac{n!}{q^n 2^{\binom{n}{2} }}  q^{-\a '},\quad \textrm{where}
  \quad  n=\lceil \log_2 L\rceil\,.  
\end{equation} 
In particular, on mesoscopic length scales $L=O(1/q^\g)$, 
$\g<1$, the constants $\a$ and $\a'$ depend only on $\g$.
For $L=\lceil d/q^\g \rceil$, with $\g\in (0,1]$ and some $d>0$,
\eqref{piscina} gives the leading order dependence of the relaxation
time  on $\theta=\log_2(1/q)$ and $\g$ as follows:
\begin{gather}
\label{piscina2}
   	   2^{F(\theta,\g) - c\theta} \leq T_{\rm{rel} }\left(\lceil d/q^\g \rceil \right)\leq 2^{F(\theta,\g)+ c' \theta}\,, \\
	\textrm{where} \quad F(\theta,\g) = \g\left(1-\g/2\right)\theta^2 + \g \theta \log_2 \theta \,,\nonumber
\end{gather}
for two positive constants $c,c'$ which are independent of $\theta$.
The bounds in \eqref{piscina} are also sufficient to show that timescale
separation does occur on mesoscopic length scales ($\g <1$).

\begin{remark}
We emphasize that \eqref{piscina} alone does not exclude some weak
form of time scale separation at
the equilibrium scale ($\g=1$) because of the presence of the unknown
exponents $\a,\a'$. Thus \eqref{eq:3} 
\emph{does not} follow
from \eqref{piscina}, the two results  required different mathematical analysis. 
\end{remark}

Heuristically the above behavior of the relaxation time can be
justified as follows.
It was shown in \cite{SE1,SE2,CDG} that, starting from the state $\sigma =
(1,1,\ldots,1,0)$ with a single vacancy at $L=2^n$, in order to flip
the rightmost spin
the system has to create at least $n$ extra vacancies. Since the
creation of a vacancy requires the overcoming
of a local \emph{energy barrier}, in a first
approximation the
non-equilibrium dynamics  of the East model for $q\ll 1$ is driven by
a non-trivial energy landscape whose characteristic activation time
$t_n$ is
of order $1/q^n$. The other main contribution in \eqref{piscina},
$n!/2^{\binom{n}{2} }$, is much more subtle and more difficult to
justify heuristically. It is an
\emph{entropic} term related, in some sense, to the number of configurations inside
$\{1,2,\dots, 2^n-1\}$ which can be reached from the state $\sigma$ using \emph{at most} $n$ vacancies. For such a
quantity, call it $V(n)$,  the following bounds were established in \cite{CDG}    
\[
c_1^n \,n!\, 2^{\binom{n}{2} }\le  V(n) \le c_2^n\, n!\, 2^{\binom{n}{2} }
\]
with $c_1,c_2$ positive constants in $(0,1)$. 

A first naive guess would be that the actual relaxation
time is the activation time $t_n$ reduced by a factor proportional to
$V(n)^{-1}$ (see \cite{CMST} for a rigorous lower bound on $T_{\rm
  rel}(L)$ based on this idea).  Notice however that the true
reduction factor in \eqref{piscina} is much smaller and equal to
$2^{\binom{n}{2} }/n!$. Thus only a tiny fraction
$\left(1/n!\right)^2$ of the
configurations reachable with $n$ vacancies really matters. What
happens is that most configurations with $n$ vacancies will return
quickly to the initial state $\s$ before removing the vacancy at $2^n$ and are therefore not
visited during the last excursion which overcomes the energy
barrier and removes the vacancy.

\noindent
\textbf{Summary.}
In this note we reported on some rigorous results about time scale
separation and dynamic heterogeneities for the low temperature East
model. We prove a strong equivalence of three characteristic
time scales which appeared in the physical literature and we
established a strong form of time scale separation up to, but
excluding, the equilibrium length scale $1/q$. A basic ingredient for these results is a novel rigorous approach
to estimating precisely the relaxation time, in which, besides the activation time across
energy barriers, key entropic contributions are accounted for. At the equilibrium length scale we prove the absence of
any form of time scale separation, including the one conjectured by Evans and Sollich which was used
to formulate their super-domain dynamics. Due to our results the
scaling limit of the stationary East model on scales $L\geq 1/q$
remains an open problem. We observe that our findings are consistent
with the following conjecture \cite{Aldous}:
\begin{conjecture}
As $q\downarrow 0$ the stationary East process in $[0,+\infty)$, after
rescaling space by $q$ and speeding up time by the relaxation time,
converges to a limit point process $X_t$ on $[0,+\infty)$ which can be
described as follows:
\begin{enumerate}[(i)]
\item At fixed time $t$, $X_t$ is a Poisson (rate $1$) process of
  particles ($\equiv 0$ spins).
\item For each $\ell>0$ and with a positive rate depending on
  $\ell$ each particle deletes all particles to its
  right up to a distance $\ell$ and replaces them by a new Poisson (rate $1$) process of particles.
\end{enumerate}  
\end{conjecture} 
At this stage we do not have a detailed understanding of the dynamics
on the equilibrium length scale, and as such there is no proposed form
for the dependence of the rate on the distance $\ell$, beyond the
fact that it is 
decreasing in $\ell$.



\begin{bibdiv}
\begin{biblist}

\bib{Aldous}{article}{
      author={Aldous, D.},
      author={Diaconis, P.},
       title={The asymmetric one-dimensional constrained Ising model: rigorous
  results},
        date={2002},
     journal={J. Stat. Phys.},
      volume={107},
      number={5-6},
       pages={945\ndash 975},
}

\bib{Berthier}{article}{
      author={Berthier, Ludovic},
       title={Dynamic heterogeneity in amorphous materials},
        date={2011},
     journal={Physics},
      volume={4},
       pages={42},
}

\bib{Bhatnagar:2007tr}{article}{
      author={Bhatnagar, N.},
      author={Caputo, P.},
      author={Tetali, P.},
      author={Vigoda, E.},
       title={{Analysis of top-swap shuffling for genome rearrangements}},
        date={2007},
     journal={Ann. of Appl. Probab.},
      volume={17},
      number={4},
       pages={1424\ndash 1445},
}

\bib{Cancrini:2006uu}{article}{
      author={Cancrini, N.},
      author={Martinelli, F.},
      author={Roberto, C.},
      author={Toninelli, C.},
       title={{Relaxation times of kinetically constrained spin models with
  glassy dynamics}},
        date={2007},
     journal={J. of Stat. Mech-Theory E.},
      volume={2007},
      number={03},
       pages={L03001},
}

\bib{CMRT}{article}{
      author={Cancrini, N.},
      author={Martinelli, F.},
      author={Roberto, C.},
      author={Toninelli, C.},
       title={Kinetically constrained spin models},
        date={2008},
     journal={Probab. Theory Rel.},
      volume={140},
      number={3-4},
       pages={459\ndash 504},
  url={http://www.ams.org/mathscinet/search/publications.html?pg1=MR&s1=MR2365481},
}

\bib{CMST}{article}{
      author={Cancrini, N.},
      author={Martinelli, F.},
      author={Schonmann, R.},
      author={Toninelli, C.},
       title={Facilitated oriented spin models: some non equilibrium results},
        date={2010},
        ISSN={0022-4715},
     journal={J. Stat. Phys.},
      volume={138},
      number={6},
       pages={1109\ndash 1123},
         url={http://dx.doi.org/10.1007/s10955-010-9923-x},
}

\bib{PietroCaputo:2012vl}{article}{
      author={Caputo, Pietro},
      author={Lubetzky, Eyal},
      author={Martinelli, Fabio},
      author={Sly, Allan},
      author={Toninelli, Fabio~Lucio},
       title={{Dynamics of 2+1 dimensional SOS surfaces above a wall: slow
  mixing induced by entropic repulsion}},
        date={2012},
      journal={arXiv:1205.6884 [math.PR]},
}

\bib{CFM}{article}{
      author={Chleboun, P.},
      author={Faggionato, A.},
      author={Martinelli, F.},
       title={Time scale separation and dynamic heterogeneity in the low
  temperature {E}ast model},
        date={2012},
        journal={arXiv:1212.2399 [math-ph]},
}

\bib{CDG}{article}{
      author={Chung, F.},
      author={Diaconis, P.},
      author={Graham, R.},
       title={Combinatorics for the {E}ast model},
        date={2001},
     journal={Adv. in Appl. Math.},
      volume={27},
      number={1},
       pages={192\ndash 206},
  url={http://www.ams.org/mathscinet/search/publications.html?pg1=MR&s1=MR1835679},
}

\bib{Derrida}{book}{
      author={Derrida, B.},
       title={Coarsening phenomena in one dimension},
      series={Lecture Notes in Physics},
   publisher={Springer},
     address={Berlin},
        date={1995},
      volume={461},
}

\bib{FMRT-cmp}{article}{
      author={Faggionato, A.},
      author={Martinelli, F.},
      author={Roberto, C.},
      author={Toninelli, C.},
       title={Aging through hierarchical coalescence in the {E}ast model},
        date={2012},
        ISSN={0010-3616},
     journal={Commun. Math. Phys.},
      volume={309},
       pages={459\ndash 495},
         url={http://dx.doi.org/10.1007/s00220-011-1376-9},
}

\bib{East-Rassegna}{article}{
      author={Faggionato, A.},
      author={Martinelli, F.},
      author={Roberto, C.},
      author={Toninelli, C.},
       title={The {E}ast model: recent results and new progresses},
        date={2012},
     journal={preprint, arXiv:1205.1607},
}

\bib{FMRT}{article}{
      author={Faggionato, A.},
      author={Martinelli, F.},
      author={Roberto, C.},
      author={Toninelli, C.},
       title={Universality in one dimensional hierarchical coalescence
  processes},
        date={2012},
     journal={Ann. Probab.},
      volume={40},
      number={4},
       pages={1377\ndash 1435},
}

\bib{FA2}{article}{
      author={Fredrickson, G.H.},
      author={Andersen, H.C.},
       title={Facilitated kinetic {I}sing models and the glass transition},
        date={1985},
     journal={J. Chem. Phys.},
      volume={83},
       pages={5822\ndash 5831},
}

\bib{GarrahanSollichToninelli}{article}{
      author={Garrahan, J.P.},
      author={Sollich, P.},
      author={Toninelli, C.},
       title={Kinetically constrained models},
     journal={in ``Dynamical heterogeneities in glasses, colloids, and granular
  media", Oxford Univ.Press, Eds.: L. Berthier, G. Biroli, J-P Bouchaud, L.
  Cipelletti and W. van Saarloos. Preprint arXiv:1009.6113},
}

\bib{JACKLE}{article}{
      author={J\"{a}ckle, J.},
      author={Eisinger, S.},
       title={A hierarchically constrained kinetic {I}sing model},
        date={1991},
     journal={Z. Phys. B: Condens. Matter},
      volume={84},
      number={1},
       pages={115\ndash 124},
}

\bib{Peres-Sly}{article}{
      author={Peres, Yuval},
      author={Sly, Allan},
       title={{Mixing of the upper triangular matrix walk}},
        date={2012-05},
     journal={Probab. Theory Rel.},
      volume={online first},
}

\bib{Ritort}{article}{
      author={Ritort, F.},
      author={Sollich, P.},
       title={Glassy dynamics of kinetically constrained models},
        date={2003},
     journal={Advances in Physics},
      volume={52},
      number={4},
       pages={219\ndash 342},
}

\bib{SE2}{article}{
      author={Sollich, P.},
      author={Evans, M.R.},
       title={Glassy time-scale divergence and anomalous coarsening in a
  kinetically constrained spin chain},
        date={1999},
     journal={Phys. Rev. Lett},
      volume={83},
       pages={3238\ndash 3241},
}

\bib{SE1}{article}{
      author={Sollich, P.},
      author={Evans, M.R.},
       title={Glassy dynamics in the asymmetrically constrained kinetic {I}sing
  chain},
        date={2003},
     journal={Phys. Rev. E},
       pages={031504},
}

\bib{Spiral}{article}{
      author={Toninelli, Cristina},
      author={Biroli, Giulio},
       title={{A new class of cellular automata with a discontinuous glass
  transition}},
        date={2008},
     journal={J. Stat. Phys.},
      volume={130},
      number={1},
       pages={83\ndash 112},
}

\end{biblist}
\end{bibdiv}

\end{document}